\documentclass[aps,prd,showpacs,floatfix,preprint,11pt]{revtex4}
\usepackage{amsmath,bm}
\usepackage{graphicx}
\usepackage{epsfig}
\usepackage{color}

\begin{document}

\title{Slowly rotating superfluid neutron stars with isospin dependent 
entrainment in a two-fluid model}
\author{Apurba Kheto and Debades Bandyopadhyay}
\affiliation{Astroparticle Physics and Cosmology Division and Centre for 
Astroparticle Physics, Saha Institute of Nuclear Physics, 1/AF Bidhannagar, 
Kolkata-700064, India}

\begin{abstract}
We investigate the slowly rotating general relativistic superfluid
neutron stars including the entrainment effect in a two-fluid model, where one 
fluid represents the superfluid neutrons and the other is the charge-neutral 
fluid called the proton fluid, made of protons 
and electrons. The equation of state and the entrainment effect between the 
superfluid neutrons and the proton fluid are computed using a relativistic mean
field (RMF) model where baryon-baryon interaction is mediated by the exchange 
of $\sigma$, $\omega$, and $\rho$ mesons and scalar self 
interactions are also included. The equations governing rotating neutron stars
in the slow rotation approximation are second 
order in rotational velocities of neutron and proton fluids. We explore the 
effects of the isospin dependent entrainment and the relative rotation between 
two fluids on the global properties of rotating superfluid neutron stars such 
as mass, shape, and the mass shedding (Kepler) limit within the RMF model with 
different parameter sets. It is observed that for the global 
properties of rotating superfluid neutron stars in particular, the Kepler limit
is modified compared with the case that does not include the contribution 
of $\rho$ mesons in the entrainment effect.  
\pacs{97.60.Jd, 47.75.+f, 95.30.Sf}
\end{abstract}
\maketitle

\section{Introduction}
The study of superfluid dynamics in neutron stars has gained momentum 
recently with the observation of fast cooling of the neutron star in 
Cassiopeia A (Cas A) 
\cite{ho}. It has been inferred that the rapid cooling in the neutron star in 
Cas A might be the outcome of neutron superfluidity in its interior\cite{page}.
The glitch phenomenon in neutron stars might also be 
strong evidence of the superfluidity in the crust and core of a neutron star
\cite{baym,itoh,and,cha}. 
There might be an interplay between the superfluidity in neutron stars and 
superfluidity studied in the laboratory. One important aspect of the 
superfluidity is the entrainment effect which was found in a mixture of
superfluid $^3$He and $^4$He in the laboratory \cite{bash}. A similar effect 
might occur 
in superfluid neutron stars when neutron and proton fluids pass through each
other. In this case, the two fluids are coupled because the momentum of one 
fluid 
carries along with it some mass current of the other fluid. This is known as 
the entrainment effect. 

The entrainment effect in superfluid neutron star matter was calculated in 
relativistic mean field (RMF) models \cite{joy,Kheto14}. Comer and
Joynt exploited the $\sigma$-$\omega$ Walecka model for this purpose. However,
neutron star matter is highly asymmetric, and the role of symmetry energy is 
very important in determining the equation of state (EoS) and the structures of 
neutron 
stars. It is expected that the symmetry energy might also influence the 
entrainment effect. Recently we investigated the entrainment effect in the 
the RMF model including $\rho$ mesons \cite{Kheto14}. We showed that the 
symmetry energy significantly affected the entrainment effect compared to 
the case without $\rho$ mesons \cite{Kheto14}. It may be worth mentioning here
that the dependence of the entrainment effect on the symmetry energy was also 
studied using polytropic equations of state \cite{prix,novak} as well as with 
relativistic Fermi liquid theory \cite{boru,haen,gus05,gus09}. 

The role of the entrainment effect in rotating neutron stars was investigated in
Newtonian as well as general relativistic formulations by different groups
\cite{prix,novak,com04}. In some of those calculations, the dependence of the 
entrainment effect on the symmetry energy was considered through the polytropic
EoS \cite{prix,novak}. However, so far, there is no calculation of rotating
neutron stars based on the 
isospin dependent entrainment effect derived from a realistic EoS.  

In this paper, we are interested in the role of isospin dependent entrainment 
on slowly rotating superfluid neutron stars. Here we adopt the two-fluid
formalism for slowly rotating superfluid neutron stars as described in 
Ref.\cite{com01}. The paper is organised in the following way. In Sec. II we 
describe the formalism for calculating the isospin dependent entrainment in a
RMF model of dense baryonic matter and the 
application of Hartle's slow rotation approximation to Einstein's field
equations for superfluid neutron stars. We discuss results in Sec. III. 
Section IV gives the summary and conclusion.      

\section{Methodology}
\subsection{The superfluid formalism}
Here we consider the superfluid formalism developed by various groups 
\cite{com01,com02,car,lan94,lan95,lan982,lan983}. The signature of the metric 
used here is the same as in Ref.\cite{joy}. 
The master function ($\Lambda$) in  the superfluid formalism is 
a function of three scalars, $n^2= - n_\mu n^\mu$, $p^2= - p_\mu p^\mu$ and  
$x^2= - n_\mu p^\mu$ which are constructed from neutron
($n^{\mu}$) and proton ($p^{\mu}$) number density currents. 
It may be noted that $-\Lambda (n^2,p^2,x^2)$ 
corresponds to the total thermodynamic energy density when neutron and proton
fluids are comoving.
The stress-energy tensor is written as 
\cite{joy,com04}
\begin{equation}
T^{\mu}_\nu = \Psi {\delta^\mu _\nu} + n^\mu \mu_\nu + p^\mu \chi_\nu
\end{equation}
and the generalized pressure is given by
\begin{equation}
    \Psi = \Lambda - n^\rho \mu_\rho - p^\rho \chi_\rho~. 
\end{equation}
The neutron and proton momentum covectors 
\begin{eqnarray}
\label{coef1}
    \mu_\nu &=& {\cal B} n_\nu + {\cal A} p_\nu  \ , \\
    \chi_\nu &=& {\cal A} n_\nu + {\cal C} p_\nu \ , 
\label{coef2}
\end{eqnarray}
are conjugate to $n^{\mu}$ and $p^{\mu}$, respectively. It is manifestly 
evident that neutron or proton momentum is a linear combination
of both number density currents. The magnitudes of which are chemical 
potentials of neutron and proton fluids, respectively \cite{joy}. 
The charge-neutral proton fluid is composed of protons and electrons, and it 
was shown that the chemical potential of the proton fluid is the sum of 
proton and electron chemical potentials \cite{lan99,prix02}. 
The master function is independent of the entrainment effect; i.e., $x^2=0$ 
when the coefficient ${\cal A}$ is zero. 
One obtains the coefficients of Eqs. ({\ref{coef1}}) and
({\ref{coef2}}) from the master function,      
\begin{equation}
    {\cal A} = -\frac {\partial \Lambda} {\partial x^2}~, 
    {\cal B} = -2\frac {\partial \Lambda} {\partial n^2}~, 
    {\cal C} = -2\frac { \partial \Lambda} {\partial p^2}~. 
\end{equation}
The field equations for neutrons and protons involve two conservation 
equations as well as two Euler equations. 

In the slow rotation approximation, the master function is written in terms
of $x^2 - n p$, which is small with respect to $n p$ \cite{joy}, 
\begin{equation}
\Lambda(n^2,p^2,x^2) = \sum_{i = 0}^{\infty} \gamma_i(n^2,p^2) 
                           \left(x^2 - n p\right)^i~. 
\label{slow}
\end{equation}

Using this form of master function, the coefficients 
$\cal A$, ${\cal A}_0^0$, etc., that determine the nonrotating background 
configuration are calculated easily \cite{joy,com04}. 

For the slow rotation approximation, we are interested in terms up to second 
order in the rotational velocities of neutrons and protons. This corresponds to
the terms proportional to $x^2 - np$ in the master function. 
It may be noted that the following combinations appearing in the field
equations are dependent on $\gamma_1$ when computed on the background 
\cite{com04}: 
\begin{eqnarray}
    {\cal A} + n \frac{\partial {\cal A}}{\partial n} + n p \frac{\partial {\cal A}}
         {\partial x^2} &=& - \gamma_1 - n \frac{\partial \gamma_1}
         {\partial n} - \sum_{i = 2}^\infty \left(\gamma_i + n 
         \frac{\partial \gamma_i}{\partial n}\right) \left(x^2 - n 
         p\right)^{i - 1} \ , \\
    && \cr
    {\cal A} + p \frac{\partial {\cal A}}{\partial p} + n p \frac{\partial {\cal A}}
         {\partial x^2} &=& - \gamma_1 - p \frac{\partial \gamma_1}
         {\partial p} - \sum_{i = 2}^\infty \left(\gamma_i + p 
         \frac{\partial \gamma_i}{\partial p}\right) \left(x^2 - n 
         p\right)^{i - 1} \ .
\end{eqnarray}

The calculation of the master function in the RMF model was described in 
detail in Refs.\cite{joy,Kheto14}. Unlike the calculation of Comer and Joynt
\cite{joy}, the role of symmetry energy on the master function and the 
entrainment
effect was considered in Ref.\cite{Kheto14}. In the latter case, the   
relativistic $\sigma$-$\omega$-$\rho$ model including scalar meson 
self-interactions \cite{bog}, was used to derive the master function 
\cite{Kheto14}. 
The Lagrangian density for nucleon-nucleon interaction 
has the form \cite{gle} 
\begin{eqnarray}\label{lag}
{\cal L}_B &=& \sum_{B=n,p} \bar\Psi_{B}\left(i\gamma_\mu{\partial^\mu} - m_B
+ g_{\sigma B} \sigma - g_{\omega B} \gamma_\mu \omega^\mu
- g_{\rho B}
\gamma_\mu{\mbox{\boldmath t}}_B \cdot
{\mbox{\boldmath $\rho$}}^\mu \right)\Psi_B\nonumber\\
&& + \frac{1}{2}\left( \partial_\mu \sigma\partial^\mu \sigma
- m_\sigma^2 \sigma^2\right) - \frac{1}{3}b m \left(g_\sigma \sigma\right)^3 
- \frac{1}{4}c  \left(g_\sigma \sigma\right)^4 \nonumber\\
&& -\frac{1}{4} \omega_{\mu\nu}\omega^{\mu\nu}
+\frac{1}{2}m_\omega^2 \omega_\mu \omega^\mu
- \frac{1}{4}{\mbox {\boldmath $\rho$}}_{\mu\nu} \cdot
{\mbox {\boldmath $\rho$}}^{\mu\nu}
+ \frac{1}{2}m_\rho^2 {\mbox {\boldmath $\rho$}}_\mu \cdot
{\mbox {\boldmath $\rho$}}^\mu ~.
\end{eqnarray}
The Dirac nucleon effective mass $m_*$ is defined as $m_*=m-<g_\sigma \sigma>$
where the nucleon mass ($m$) is taken as the average of bare neutron 
($m_n$) and proton ($m_p$) masses. 
The frame in which neutrons have zero 
spatial momentum and protons have a wave vector $k_{\mu}=(k_0,0,0,K)$ 
\cite{joy} is chosen to solve the equations of motion for meson fields 
in the mean field approximation \cite{gle}. 

The master function, generalized pressure, and chemical potentials of neutron 
and proton fluids in the limit $K \rightarrow 0$ are given by 
\begin{eqnarray}
    \left.\Lambda\right|_0 &=&  - {c_\omega^2\over {18 \pi^4}}\left(k_n^3+k_p^3\right)^2-{c_\rho^2\over {72 \pi^4}}\left(k_p^3-k_n^3\right)^2- {1 \over 4 \pi^2} \left(k_n^3 
             \sqrt{k_n^2 + \left.m^2_*\right|_0} + 
             k_p^3 \sqrt{k_p^2 + \left.m^2_*\right|_0}
             \right)   \cr
             && \cr
             &&- {1 \over 
             4} c_\sigma^{- 2} \left[\left(2 m - \left.m_*\right|_0\right) 
             \left(m - \left.m_*\right|_0\right)+\left.m_*\right|_0\left( b m c_\sigma^2\left(m-\left.m_*\right|_0\right)^2 
          + c  c_\sigma^2\left(m-\left.m_*\right|_0\right)^3\right)\right]\nonumber \\
          &&-\frac{1}{3} b m \left(m-\left.m_*\right|_0\right)^3\--\frac{1}{4} c  \left(m-\left.m_*\right|_0\right)^4-{1 \over 8 \pi^2} \left( k_p \left[2 k_p^2 + m_e^2\right] \sqrt{k_p^2 + m^2_e}\right.\nonumber \\
             &&
             \left. - m^4_e {\rm ln}\left[
             {k_p + \sqrt{k_p^2 + m^2_e} \over m_e}\right]\right) 
              \ , \\
             && \cr
    \left.\mu\right|_0 &=&-\frac{\pi^2}{k_n^2} \left.\frac{\partial \Lambda}{\partial k_n}\right|_0={c_\omega^2\over {3 \pi^2}}\left(k_n^3+k_p^3\right)
 - {c_\rho^2\over {12 \pi^2}}\left(k_p^3-k_n^3\right)  + \sqrt{k_n^2 + 
             \left.m^2_*\right|_0} \ , \\ 
             && \cr
    \left.\chi\right|_0 &=&-\frac{\pi^2}{k_p^2} \left.\frac{\partial \Lambda}{\partial k_p}\right|_0= {c_\omega^2\over {3 \pi^2}}\left(k_n^3+k_p^3\right)
+ {c_\rho^2\over {12 \pi^2}}\left(k_p^3-k_n^3\right) + \sqrt{k_p^2 + 
             \left.m^2_*\right|_0} + \sqrt{k_p^2 + m_e^2} \  , \\ 
             && \cr
    \left.\Psi\right|_0 &=& \left.\Lambda\right|_0 + {1 \over 3 \pi^2}
             \left(\left.\mu\right|_0 k_{n}^3 + 
             \left.\chi\right|_0 k_{p}^3\right)~,
\end{eqnarray}
where the subscript "0" stands for quantities calculated in the limit 
$K \rightarrow 0$, $c_{\sigma}^{2}=(g_\sigma/m_\sigma)^2$ , 
$c_{\omega}^2=(g_\omega/m_\omega)^2$, and  $c_{\rho}^2=(g_\rho/m_\rho)^2$, 
and
\begin{eqnarray}
    \left.m_*\right|_0 &=& m_*(k_n,k_p,0) \cr
         && \cr
        &=& m - \left.m_*\right|_0 {c_\sigma^2 \over 2 \pi^2}  
            \left(k_n \sqrt{k_n^2 + \left.m^2_*\right|_0} + k_p 
            \sqrt{k_p^2 + \left.m^2_*\right|_0} + {1 \over 2} 
            \left.m_*^2\right|_0 {\rm ln} \left[{- k_n + 
            \sqrt{k_n^2 + \left.m^2_*\right|_0} \over k_n + 
            \sqrt{k_n^2 + \left.m^2_*\right|_0}}\right] \right. \cr
         && \cr
         && + {1 \over 2} \left.\left.m_*^2\right|_0 {\rm ln} \left[{- 
            k_p + \sqrt{k_p^2 + \left.m^2_*\right|_0} \over k_p + 
            \sqrt{k_p^2 + \left.m^2_*\right|_0}}\right]\right) \ + b  m  c_\sigma^2\left(m - m_*\right)^2 +   c  c_\sigma^2\left(m - m_*\right)^3~. 
            \label{mstar}
\end{eqnarray}
It is to be noted here that electrons are treated as noninteracting 
relativistic particles and are included in the calculation of the master '
function and generalised pressure.
The values of the various coefficients ${\cal A}|_0$, ${\cal B}|_0$, 
${\cal C}|_0$, ${\cal A}_0^0|_0$, ${\cal B}_0^0|_0$ and ${\cal C}_0^0|_0$ that 
appear in the field equations are provided in the Appendix. 

\subsection{Slowly rotating superfluid neutron stars}

Andersson and Comer \cite{com01} extended Hartle's slow rotation formalism for
the single fluid \cite{Hartle67} to the case of the two-fluid model in order to 
describe superfluid 
neutron stars. They considered that the superfluid neutron and the proton fluid
are rotating with different rotational velocities. However, they did not 
include the entrainment effect in their calculation. Here we adopt the 
two-fluid formalism of Andersson and Comer as described by 
Refs. \cite{com01,com04} to study 
stationary, axisymmetric, and asymptotically flat configurations. Furthermore we
introduce the isospin dependent entrainment in this calculation.  
In the slow rotation 
approximation, rotational velocities of neutron ($\Omega_n$) and proton 
($\Omega_p$) fluids are considered as small so that inequalities 
$\Omega_n R << c$ and $\Omega_p R << c$ are satisfied, where $c$ is the speed 
of light.
The slow rotation acts as the perturbation on nonrotating configurations.
We retain terms up to second order in the angular velocities of neutron and 
proton fluids 
in field equations in the slow rotation approximation.
The metric used here has the following structure \cite{Hartle67,com01,com04}: 
\begin{equation}
{\rm g}_{\mu\nu}{\rm d}x^{\mu} {\rm d}x^{\nu} = - (N^2 -{\rm sin}^2{\theta} K[N^{\phi}]^2){\rm d}t^2 
             + V {\rm d}{\tilde r}^2 
             - 2KN^{\phi}{\rm sin}^2{\theta} {\rm d}t {\rm d}{\phi}
             + K \left({\rm d}\theta^2 + {\rm sin}^2\theta 
             {\rm d}\phi^2\right) \ . \label{bgmet}
\end{equation}
The equations relevant for the metric variables in the two-fluid model and
the slow rotation approximation are same as those of Hartle's 
single-fluid model and the metric functions are expanded in powers of 
angular velocities \cite{com01,com04,Hartle67}, 
 \begin{eqnarray}
    N &=& e^{\nu(\tilde{r})/2} \left(1 + h(\tilde {r},\theta)\right) 
          \ , \cr
       && \cr 
    V &=& e^{\lambda(\tilde {r})} \left(1 + 2 v(\tilde {r},\theta)\right) 
          \ , \cr 
       && \cr
    K &=& \tilde {r}^2 (1 + 2 k(\tilde {r},\theta)) \ , \cr
       && \cr
    N^{\phi} &=& \omega(\tilde {r}) \ , 
\end{eqnarray}
where $\omega$ is a first order quantity in angular velocities, and 
$h$, $v$, and $k$ are second order quantities.
Further $h$, $v$, 
and $k$ are decomposed into  $\ell = 0$ and $\ell = 2$ terms after 
expanding those in spherical harmonics, 
\begin{eqnarray}
    h &=& h_0(\tilde {r}) + h_2(\tilde {r}) P_2({\cos}\theta) \ , \cr
       && \cr
    v &=& v_0(\tilde {r}) + v_2(\tilde {r}) P_2({\cos}\theta) \ , \cr
       && \cr
    k &=& k_2(\tilde {r}) P_2({\cos}\theta) \ ,
\end{eqnarray}
where $P_2({\cos}\theta) = (3 {\cos}^2\theta - 1)/2$. 

Similarly, neutron ($n$) and proton ($p$) number densities are expanded as 
\begin{equation}
    n = n_0 (\tilde{r}) \left(1 + \eta(\tilde{r},\theta)\right) \qquad , \qquad 
    p = p_0 (\tilde {r}) \left(1 + \Phi(\tilde {r},\theta) 
         \right) \ ,
\end{equation}
where terms $\eta$ and $\Phi$ are of ${\cal O}(\Omega^2_{n,p})$, 

\begin{equation}
    \eta = \eta_0(\tilde {r}) + \eta_2(\tilde {r}) P_2({\cos}\theta) 
             \quad , \quad
    \Phi = \Phi_0(\tilde {r}) + \Phi_2(\tilde {r}) P_2({\cos}\theta)  
             \ .
\end{equation}

A coordinate transformation $\tilde{r} \to r + \xi(r,\theta)$ is introduced 
such that $\Lambda(\tilde{r}(r,\theta),\theta)=\Lambda_0(r)$ \cite{com01}. 
Here the $\xi$ coordinate is also expanded in spherical harmonics as
$\xi = \xi_0(r) + \xi_2(r) P_2({\cos}\theta)$. 

With this prescription of the slow rotation approximation for metric functions 
as well as neutron and proton densities along with the coordinate 
transformation,
the fluid and Einstein field equations are reduced to four sets of equations.
The first set of equations corresponds to nonrotating  background 
configurations that are obtained from the solutions of two
background metric components $\lambda$ and $\nu$ \cite{com01,com04}. Those are 
given in terms of coefficients of fluid equations,
\begin{equation}
    \left. A^0_0\right|_0  p_0^{\prime} + \left. B^0_0\right|_0 
     n_0^{\prime} + {1 \over 2} \left.\mu\right|_0 \nu^{\prime} = 0 
    \quad , \quad
    \left. C^0_0\right|_0  p_0^{\prime} + \left. A^0_0\right|_0 
     n_0^{\prime} + {1 \over 2} \left.\chi\right|_0 \nu^{\prime} = 0 
    \ , \label{comer}
\end{equation}
where prime denotes differentiation with respect to $\tilde {r}$ and 
$\left.A_0^0\right|_0$, $\left.B_0^0\right|_0$, and 
$\left.C_0^0\right|_0$ coefficients are obtained from the master 
function and are taken from Ref.\cite{Kheto14}.
The regularity condition demands that $\lambda$, 
$\lambda^{\prime}$, $\nu^{\prime}$, $n_0^{\prime}$ and 
$p_0^{\prime}$ vanish at the origin. The total  mass of this configuration is 
\begin{equation}
    M = - 4 \pi \int_0^R \Lambda_0(\tilde {r}) \tilde {r}^2 {\rm d} \tilde {r} \ . 
\end{equation}

Next, the frame dragging $\omega(r)$, which is first order in angular velocities
of neutron and proton fluids, is obtained from the following equation 
\cite{com04,Hartle67}
\begin{equation}
    \frac{1}{r^4}\frac{d}{dr} \left(r^4 e^{- (\lambda + \nu)/2} 
    \frac{d \tilde{L}_{n}}{dr}\right) - 16 \pi e^{(\lambda - 
    \nu)/2} \left(\Psi_0 - \Lambda_0\right) \tilde{L}_{n} 
    = 16 \pi e^{(\lambda - \nu)/2} \chi_0 p_0 
    \left(\Omega_{n} - \Omega_{p}\right) \ . \label{framedrag}
\end{equation}
This equation has the same structure as that of the single fluid 
except for the 
nonzero term on the right-hand side \cite{Hartle67}.
Here we define $\tilde {L_n}=\omega - \Omega_n$ and $\tilde {L_p}=\omega 
- \Omega_p$, which represent the rotational frequencies as measured by a distant
observer.
The  boundary condition implies that the interior solution of $\omega(r)$
matches with the vacuum solution 
\begin{equation}
    \tilde{L}_n(R) = - \Omega_n + \frac{2 J}{R^3} \ , \label{b.c.1}
\end{equation}
where $J$ is the total angular momentum of the system.
The derivative of the solution is also continuous at the surface \cite{com04}.

The neutron and proton angular momenta, $J_n$ 
and $J_p$, respectively, are given by \cite{com01}
\begin{equation}
J_{n} = - \frac{8 \pi}{3} \int_0^R {\rm d}r r^4 e^{(\lambda - \nu)/2}
\left[\mu_0 n_0 \tilde{L}_{n} + A_0
n_0 p_0 \left(\Omega_{n} - \Omega_{p}\right)\right]
\end{equation}
and 

\begin{equation}
J_{p} = - \frac{8 \pi}{3} \int_0^R {\rm d}r r^4 e^{(\lambda - \nu)/2}
         \left[\chi_0 p_0 \tilde{L}_{p} + A_0
         n_0 p_0 \left(\Omega_{p} - \Omega_{n}\right)
         \right] \ .
\end{equation}
The total angular momentum  $J$ is equal to $J_n + J_p$.

The last two sets of equations are ${\cal O}(\Omega^2_{n,p})$ equations. One 
can obtain $\xi_0$, $\eta_0$, $\Phi_0$, $h_0$, and $v_0$ from 
$\ell = 0$ second-order equations, on the other hand, $\xi_2$, $\eta_2$, 
$\Phi_2$, $h_2$, $v_2$, and $k_2$ follow from $\ell = 2$ second-order equations.
A detailed discussion of $\ell =0$ and $\ell =2$ second order equations and
numerical techniques to solve those equations can be found in 
Refs.\cite{com01,com04,Hartle67}. After obtaining a complete solution in the
slow rotation approximation, one can calculate the quadruple moment of the
configuration and the rotationally induced change of mass as described in
Refs.\cite{com01,com04}. Furthermore,   
the Kepler frequency of the  slowly rotating superfluid neutron star is 
obtained following the prescription of Andersson and Comer\cite{com01}, 
\begin{equation}
     \Omega_K = \sqrt{\frac{M}{R^3}} - \frac{\hat{J} \Omega_p}{R^3} + 
                \sqrt{\frac{M}{R^3}} \left\{\frac{\delta \hat{M}}{2M} + 
                \frac{(R + 3 M) (3 R - 2 M)}{4 R^4 M^2} \hat{J}^2 -
                \frac{3}{4} \frac{2 \hat{\xi}_0 - \hat{\xi}_2}{R} + 
                \alpha \hat{A} \right\} \Omega_p^2~, 
\label{Kep}
\end{equation}
where scaling of $J = \hat{J} \Omega_p$, 
$\delta M = \delta \hat{M} \Omega_p^2$, 
$\xi_0 = \hat{\xi}_0 \Omega_p^2$, and $\xi_2 = \hat{\xi_2} \Omega_p^2$ with 
$\Omega_p$ is made and
\begin{equation}
    \alpha = \frac{3 (R^3 - 2 M^3)}{4 M^3} \log \left(1 - \frac{2 M}{R} 
             \right) + \frac{3 R^4 - 3 R^3 M - 2 R^2 M^2 - 8 R M^3 + 6 
             M^4}{2 R M^2(R - 2 M)} \ .
\end{equation}

It is to be noted that the expression for the Kepler frequency in 
Eq. (\ref{Kep}) differs from that of Eq. (77) of Ref.\cite{com01}. This 
difference originates from the factor at the beginning of the third term within
the second bracket and the term involving $\hat{\xi}_0$ and $\hat{\xi}_2$ in 
both equations. We discuss this issue further in the next section.
 
It is worth mentioning here that the model based on the slow rotation 
approximation is applicable for the fastest observed pulsar 
as noted by others \cite{prix,com01}. However, this approximation breaks down
near the Kepler limit \cite{prix}. 

\section{Results and Discussion}

Now we discuss the results of slowly rotating superfluid neutron stars. 
Nonrotating background configurations are obtained by solving 
Eq. (\ref{comer}). In 
this context, we exploit the RMF EoS which includes the isospin dependent 
entrainment effect \cite{Kheto14}. We use GL and NL3 parameter sets in this 
calculation, 
both of which are listed in Table I. Central neutron number density
is an essential input for the calculation of the background configurations. The 
proton number density in the 
background model is no longer a free parameter because the chemical equilibrium 
is imposed at the centre of the star, i.e., $\mu|_0 = \chi|_0$ \cite{lan99}. 
The chemical equilibrium is established when both fluids are corotating. 
However, the chemical equilibrium does not hold good for 
different rotation rates of neutron and proton fluids \cite{prix,novak}. 
Masses and radii corresponding to two nonrotating configurations are also
recorded in Table I. The chosen background configurations are just below 
their maximum masses \cite{Kheto14}. Furthermore, we consider $\eta_0 (0) = 0$ 
and, consequently, $\Phi_0 (0) = 0$ in all cases. 

As soon as we know the background configuration, we can calculate the 
frame-dragging frequency from Eq.(\ref{framedrag}). As we are dealing with the 
two-fluid system, the central value of $\tilde {L}_n$ and relative rotation 
rate $\Omega_n/\Omega_p$ are needed to solve Eq.(\ref{framedrag}) 
\cite{com01}.
A rescaled Eq.(\ref{framedrag}) with the definition of 
$\hat L_n(r) = \tilde {L}_n / \Omega_p$ is solved to determine the 
frame-dragging frequency for different values of $\Omega_p$ using a fixed 
relative rotation rate. The boundary condition of the problem demands that   
the interior solution matches with the known vacuum solution given by 
Eq.(\ref{b.c.1}). 
The frame-dragging frequency, $\frac{\omega(r)}{\Omega_p}$, is
plotted as a function of radial distance ($r/R$) in Fig. 1  for three 
different 
relative rotation rates. The left panel denotes the GL parameter set and
the right panel represents the NL3 set. The frame-dragging frequency decreases 
monotonically from the centre to the surface of the star for three
relative rotation rates in both panels. This feature of the 
frame-dragging frequency is quite similar to the standard single-fluid result
\cite{Hartle68}. Further it is noted that the frame-dragging frequency is 
always higher for larger values of relative rotation rate.
 
Now we  discuss numerical solutions of different metric functions of the 
superfluid neutron star in the slow rotation approximation . 
First we  solve the $\ell=0$ equations and determine $\xi_0$, $\eta_0$, 
$\Phi_0$, $h_0$ and $v_0$ following the procedure laid down by Andersson and 
Comer {\cite{com01}}. Metric functions $h_0$ and $v_0$ match with the vacuum 
solutions at the surface. The metric function $v_0(r)$  as a function of radial
distance is displayed in Fig. 2 for three different relative rotation rates. 
The left panel shows the results of the GL set and the right panel 
corresponds to those of the NL3 set. It is noted that the metric function 
$v_0$ increases monotonically to the surface and matches smoothly with
the exterior solution. For the NL3 set, the value of this metric function
at the surface is always higher than that of the GL set. 
A new metric function $m_0$ is defined in terms of $v_0$ and $\lambda$ as 
$m_0 = r v_0/\exp(\lambda)$. 
The radial profile of $m_0$, which merges with the exterior solution at the 
surface, is shown in Fig. 3 for different relative rotation rates. 

We solve the $\ell=2$ equations in a similar way to that used for solutions
of $\ell=0$ equations \cite{com01}. A new variable, ${\bar k}=k_2 + h_2$, is 
introduced to solve two coupled first-order equations in $h_2$ and $k_2$ 
\cite{Hartle67}.
This leads to two coupled differential equations in $\bar k$ and $h_2$, which 
are
solved using the method described by Hartle \cite{Hartle67}. In Fig. 4 , 
the metric functions $h_0$ and $h_2$ are plotted as a function of radial 
distance for different relative rotation rates. 
The results of the GL and NL3 sets are shown in the left and right panels,
respectively. In both panels, the lower three curves denote the metric function
$h_0$ and the upper three curves imply the metric function $h_2$. 
Figure 5 shows
the radial profiles of $\xi_0(r)$ (upper curves) and $\xi_2(r)$ (lower
curves), and in Fig. 6, we have 
${\bar k}=k_2 + h_2$ versus $r$  for the GL (left panel) and NL3 (right panel) 
sets and 
three different relative rotation rates. In Fig. 5, the magnitude of 
$\xi_2(r)$ at 
the surface in the right panel is quite large with respect to that of the left 
panel when the relative rotation rate is larger than 1 and this function
is directly related to the deformation of the star due to rotation.
   
Figure 7 exhibits the variation of rotationally induced corrections to the 
neutron number density $n_0 \eta_0$ (three upper curves) and $ n_0 \eta_2$ 
(three lower curves) with radius. Similarly, 
Fig. 8 represents the variation of rotationally induced corrections to the
proton number density $p_0 \Phi_0$ (three upper curves) and $ p_0 \Phi_2$ 
(three lower curves) with radius. In both cases, the left panel denotes the 
results of the GL set and the right panel corresponds to those of the NL3 
set. We explore the role of symmetry energy on the rotationally induced
corrections to the proton number density by comparing two cases with 
(left panel) and without (right panel) $\rho$ mesons for the GL set in 
Fig. 9. For the case without $\rho$ mesons, we consider a nonrotating
configuration that is just below the maximum mass neutron star. The mass and 
radius of this neutron star is 2.33 M$_{\odot}$ and 10.96 km. It is noted 
that the corrections to the proton number density are significantly modified
in the presence  of $\rho$ mesons. 

The deformation of a rotating star is obtained in terms of the ratio 
of the polar and equatorial radii. For the slowly rotating star, this is 
given by $\frac{R_p}{R_e}\approx 1 + \frac{3 \xi_2(R)}{2R}$.
The ratio of polar to equatorial radii  as a function of relative 
rotation rate is plotted in Fig. 10 for the GL (solid line) and NL3 
(dashed line) sets. We consider the proton rotation rate to be equal to that
of the fastest rotating pulsar having spin frequency 716 Hz \cite{hess}. The 
nonrotating situation is achieved when the relative rotation rate approaches 
zero.  Furthermore we find that the 
rotationally induced deformation of the star is larger for the NL3 case than 
the GL case. This deformation increases with increasing relative rotation 
rate.
  
As neutron and proton fluids may rotate at different rates, one of them
extends beyond the other at the equator. The Kepler limit is obtained from the
rotation rate of the outer fluid.
To determine the mass-shedding (Kepler) limit we have to solve the 
quadratic equation (\ref{Kep}) for $\Omega_p$. When $\Omega_n > \Omega_p$ the
Kepler frequency is determined by the neutrons; for 
$\Omega_p > \Omega_n $, the Kepler frequency is determined by the protons.
We calculate the Kepler limit in the RMF model including $\rho$ mesons 
using the GL and NL3 parameter sets for the background configurations of 
Table I.
The mass-shedding (Kepler) limit $\Omega_K$ as a function of relative
rotation rate is plotted in Fig. 11 for the GL set (left panel) and the 
NL3 set (right panel). We use the radial profiles of the entrainment 
effect in this calculation of Kepler frequency. 
The results are qualitatively similar to 
the previous investigation by Prix and collaborators 
\cite{prix} though the authors in the that case used some constant values of
entrainment. However, our results are different from those of Comer 
\cite{com04}. For $\Omega_n > \Omega_p$, the 
Kepler frequency (solid square) approaches a constant value with increasing 
$\Omega_n$. 
When  $\Omega_n/\Omega_p < 1$ , the Kepler frequency (solid
circle) 
monotonically increases with decreasing relative rotation rate, as evident
from Fig. 11, 
whereas the opposite scenario was found in the work of Comer \cite{com04}. 
On the other hand, Prix {\it et al.} \cite{prix} found that the Kepler limit 
increased monotonically as the relative rotation rate 
decreased. This is quite similar to our results. The difference between our
results and those of Comer \cite{com04} may be due to different expressions for
$\Omega_K$ that we have discussed in connection with Eq.(\ref{Kep}) in Sec.
II. Furthermore, Comer \cite{com04} calculated the entrainment using the 
the equation of state obtained in the relativistic $\sigma$-$\omega$ model. 
Without $\rho$ mesons, the effects of symmetry energy on the entrainment 
was absent. On the other hand,  we exploit an isospin dependent entrainment  
effect calculated in the $\sigma$-$\omega$-$\rho$ RMF model for the
determination of the Kepler limit \cite{Kheto14}. We compare the Kepler limit 
calculated in the RMF model with and without $\rho$ mesons for the GL
set in Fig. 12. 
In both cases, we consider nonrotating configurations that are just below
their maximum masses, as noted in Table I and discussed in connection with 
Fig. 7. The 
solid line denotes the calculation without $\rho$ mesons and the dashed line
represents the case with $\rho$ mesons. Furthermore, solid squares and circles
correspond to allowed rotation rates of neutron and proton fluids,
respectively. It is noted that the two results 
differ, as is evident from the highlighted part of Fig. 12. 
 
\section{Summary and conclusions}
We have studied the role of the isospin dependent entrainment and the relative
rotation rates of neutron and proton fluids on the global properties of slowly
rotating superfluid neutron stars such as the structures and the Kepler limit
in the two-fluid formalism. The two-fluid formalism of Andersson and Comer
\cite{com04} is adopted in our work. The effects of symmetry energy on the EoS 
and 
entrainment are studied using the $\sigma$-$\omega$-$\rho$ RMF model. The 
symmetry
energy significantly influences the rotationally induced corrections to 
the proton number
density. It is found that the Kepler limit obtained with the isospin dependent
entrainment effect is lower than that of the case when the isospin term is 
neglected in the entrainment effect. The behaviour of the Kepler limit 
as a function of the relative rotation rate in our case is qualitatively similar
to the results of Prix {\it et al.} \cite{prix} obtained using the polytropic 
EoS.
The calculation of slowly rotating superfluid neutron stars including the
isospin dependent entrainment effect in a realistic EoS is the first
of its kind. 
\section{Appendix}
The values of some useful matter coefficients (see \cite{Kheto14})that are the 
inputs of field equations are the following
\begin{eqnarray}
{\cal A}|_0 &=& c_{\omega}^2-\frac{1}{4} c_{\rho}^2 + {c^2_{\omega} \over 5 
        \left.\mu^2\right|_0} \left(2 k_p^2 {\sqrt{k_n^2 + 
        \left.m^2_*\right|_0} \over \sqrt{k_p^2 + 
        \left.m^2_*\right|_0}} + {c^2_{\omega} \over 3 \pi^2} 
        \left[{k_n^2 k_p^3 \over \sqrt{k_n^2 + 
        \left.m^2_*\right|_0}} + {k_p^2 k_n^3 \over \sqrt{k_p^2 + 
        \left.m^2_*\right|_0} }\right]\right)\cr
         && \cr
         &&
        +{c^2_{\rho} \over 20 
        \left.\mu^2\right|_0} \left(2 k_p^2 {\sqrt{k_n^2 + 
        \left.m^2_*\right|_0} \over \sqrt{k_p^2 + 
        \left.m^2_*\right|_0}} + {c^2_{\rho} \over 12 \pi^2} 
        \left[{k_n^2 k_p^3 \over \sqrt{k_n^2 + 
        \left.m^2_*\right|_0}} + {k_p^2 k_n^3 \over \sqrt{k_p^2 + 
        \left.m^2_*\right|_0} }\right]\right)\cr
         && \cr 
         &&
         -{c^2_{\rho}c^2_{\omega} \over 30\left.\mu^2\right|_0 \pi^2} 
        \left[{k_n^2 k_p^3 \over \sqrt{k_n^2 + 
        \left.m^2_*\right|_0}} - {k_p^2 k_n^3 \over \sqrt{k_p^2 + 
        \left.m^2_*\right|_0} }\right]  + {3 \pi^2 k_p^2 
        \over 5 \left.\mu^2\right|_0 k_n^3} {k_n^2 + 
        \left.m^2_*\right|_0 \over \sqrt{k_p^2 + 
        \left.m^2_*\right|_0}} \ , \\
        && \cr
{\cal B}|_0 &=& {3 \pi^2 \left.\mu\right|_0 \over k_n^3} - 
        c_{\omega}^2 {k_p^3 \over k_n^3}+ \frac{1}{4}c_{\rho}^2 {k_p^3 \over k_n^3} - {c^2_{\omega} k_p^3 
        \over 5 \left.\mu^2\right|_0 k_n^3} \left(2 k_p^2 
        {\sqrt{k_n^2 + \left.m^2_*\right|_0} \over \sqrt{k_p^2 + 
        \left.m^2_*\right|_0}} + {c^2_{\omega} \over 3 \pi^2} 
        \left[{k_n^2 k_p^3 \over \sqrt{k_n^2 + 
        \left.m^2_*\right|_0}} + {k_p^2 k_n^3 \over \sqrt{k_p^2 + 
        \left.m^2_*\right|_0} }\right]\right)  \cr
        && \cr
        && - {c^2_{\rho} k_p^3 
        \over 20 \left.\mu^2\right|_0 k_n^3} \left(2 k_p^2 
        {\sqrt{k_n^2 + \left.m^2_*\right|_0} \over \sqrt{k_p^2 + 
        \left.m^2_*\right|_0}} + {c^2_{\rho} \over 12 \pi^2} 
        \left[{k_n^2 k_p^3 \over \sqrt{k_n^2 + 
        \left.m^2_*\right|_0}} + {k_p^2 k_n^3 \over \sqrt{k_p^2 + 
        \left.m^2_*\right|_0} }\right]\right)  \cr
        && \cr
        && +{c^2_{\rho} c^2_{\omega}k^3_p\over 30 \pi^2\left.\mu^2\right|_0 k_n^3} 
        \left[{k_n^2 k_p^3 \over \sqrt{k_n^2 + 
        \left.m^2_*\right|_0}} - {k_p^2 k_n^3 \over \sqrt{k_p^2 + 
        \left.m^2_*\right|_0} }\right]-{3 \pi^2 k_p^5 \over 5 \left.\mu^2\right|_0 k_n^6} 
        {k_n^2 + \left.m^2_*\right|_0 \over \sqrt{k_p^2 + 
        \left.m^2_*\right|_0}}\,\label{b00} \ , \\
        && \cr
{\cal C}|_0 &=& {3 \pi^2 \left.\chi\right|_0 \over k_p^3}+ \frac{1}{4}c_{\rho}^2 {k_n^3 \over k_p^3} - 
        c_{\omega}^2 {k_n^3 \over k_p^3} - {c^2_{\omega} k_n^3 
        \over 5 \left.\mu^2\right|_0 k_p^3} \left(2 k_p^2 
        {\sqrt{k_n^2 + \left.m^2_*\right|_0} \over \sqrt{k_p^2 + 
        \left.m^2_*\right|_0}} + {c^2_{\omega} \over 3 \pi^2} 
        \left[{k_n^2 k_p^3 \over \sqrt{k_n^2 + 
        \left.m^2_*\right|_0}} + {k_p^2 k_n^3 \over \sqrt{k_p^2 + 
        \left.m^2_*\right|_0} }\right]\right)  \cr
        && \cr
        && - {c^2_{\rho} k_n^3 
        \over 20 \left.\mu^2\right|_0 k_p^3} \left(2 k_p^2 
        {\sqrt{k_n^2 + \left.m^2_*\right|_0} \over \sqrt{k_p^2 + 
        \left.m^2_*\right|_0}} + {c^2_{\rho} \over 12 \pi^2} 
        \left[{k_n^2 k_p^3 \over \sqrt{k_n^2 + 
        \left.m^2_*\right|_0}} + {k_p^2 k_n^3 \over \sqrt{k_p^2 + 
        \left.m^2_*\right|_0} }\right]\right)\cr
        && \cr
        && +{c^2_{\rho} c^2_{\omega}k_n^3\over 30 \pi^2\left.\mu^2\right|_0 k_p^3} 
        \left[{k_n^2 k_p^3 \over \sqrt{k_n^2 + 
        \left.m^2_*\right|_0}} - {k_p^2 k_n^3 \over \sqrt{k_p^2 + 
        \left.m^2_*\right|_0} }\right]-{3 \pi^2 \over 5 \left.\mu^2\right|_0 k_p} 
        {k_n^2 + \left.m^2_*\right|_0 \over \sqrt{k_p^2 + 
        \left.m^2_*\right|_0}}~.
\label{coo} 
\end{eqnarray}

\begin{eqnarray}
{{\cal A}_0^0}|_0 &=-& \frac{\pi^4}{k_p^2k_n^2} \left.\frac{\partial^2 \Lambda}{\partial k_p\partial k_n}\right|_0
        = c_\omega^2 - \frac{c_\rho^2}{4}+ {\pi^2 \over k^2_p} { 
        \left.m_*\right|_0 \left.{\partial m_* \over \partial k_p}
        \right|_0 \over \sqrt{k^2_n + \left.m^2_*\right|_0}}\ , \\
        && \cr
{{\cal B}_0^0}|_0 &=& \frac{\pi^4}{k_n^5} \left(\left.2\frac{\partial \Lambda}{\partial k_n}\right|_0-k_n\left.\frac{\partial^2 \Lambda}{\partial k_n^2}\right|_0\right) 
        = c_\omega^2 + \frac{c_\rho^2}{4} + {\pi^2 \over k^2_n} {k_n + 
        \left.m_*\right|_0 \left.{\partial m_* \over \partial k_n}
        \right|_0 \over \sqrt{k^2_n + \left.m^2_*\right|_0}} \ , \\
        && \cr
{{\cal C}_0^0}|_0 &=& \frac{\pi^4}{k_p^5} \left(\left.2\frac{\partial \Lambda}{\partial k_p}\right|_0-k_p\left.\frac{\partial^2 \Lambda}{\partial k_p^2}\right|_0\right)
        = c_\omega^2 +  \frac{c_\rho^2}{4} + {\pi^2 \over k^2_p} {k_p + 
        \left.m_*\right|_0 \left.{\partial m_* \over \partial k_p}
        \right|_0 \over \sqrt{k^2_p + \left.m^2_*\right|_0}} + 
        {\pi^2 \over k_p} {1 \over \sqrt{k^2_p + m^2_e}}~.
\end{eqnarray}
where
\begin{eqnarray}
     \left.{\partial m_* \over \partial k_n}\right|_0 &=& - 
          {c_\sigma^2 \over \pi^2} {\left.m_*\right|_0 k_n^2 
          \over \sqrt{k_n^2 + \left.m^2_*\right|_0}} \left({3 m - 2 
          \left.m_*\right|_0 +3 b m c_\sigma^2\left(m-\left.m_*\right|_0\right)^2 
          +3 c  c_\sigma^2\left(m-\left.m_*\right|_0\right)^3\over\left.m_*\right|_0}\right.\nonumber \\ 
          &&\left.- {c_\sigma^2 
          \over \pi^2} \left[{k_n^3 \over \sqrt{k_n^2 + \left.m^2_*
          \right|_0}} + {k_p^3 \over \sqrt{k_p^2 + 
          \left.m^2_*\right|_0}}\right]+2 b m c_\sigma^2\left(m-\left.m_*\right|_0\right) 
          +3 c  c_\sigma^2\left(m-\left.m_*\right|_0\right)^2\right)^{- 1}~, 
          \\
          && \cr
     \left.{\partial m_* \over \partial k_p}\right|_0 &=& - 
          {c_\sigma^2 \over \pi^2} {\left.m_*\right|_0 k_p^2 
          \over \sqrt{k_p^2 + \left.m^2_*\right|_0}} \left({3 m - 2 
          \left.m_*\right|_0 +3 b m c_\sigma^2\left(m-\left.m_*\right|_0\right)^2 
          +3 c  c_\sigma^2\left(m-\left.m_*\right|_0\right)^3\over\left.m_*\right|_0}\right.\nonumber \\ 
          &&\left.- {c_\sigma^2 
          \over \pi^2} \left[{k_n^3 \over \sqrt{k_n^2 + \left.m^2_*
          \right|_0}} + {k_p^3 \over \sqrt{k_p^2 + 
          \left.m^2_*\right|_0}}\right]+2 b m c_\sigma^2\left(m-\left.m_*\right|_0\right) 
          +3 c  c_\sigma^2\left(m-\left.m_*\right|_0\right)^2\right)^{- 1} .
\end{eqnarray}

\newpage
\begin{table}[h]
\caption{Nucleon-meson coupling constants corresponding to
the GL and NL3 sets are taken from Refs.\cite{gle,horo}. 
The coupling constants are obtained by reproducing the saturation properties
of symmetric nuclear matter as detailed in Ref.\cite{Kheto14} .
All the coupling constants are in fm$^{2}$, except $b$ and $c$ which are 
dimensionless. The background nonrotating configurations as computed by Kheto 
and Bandyopadhyay \cite{Kheto14} are used here . 
The central neutron wave number $k_n(0)$ is given by fm$^{-1}$.
The mass$(M)$ and radius $(R)$ are in units of $M_\odot$ and km, respectively.   }

\begin{center}
\vspace{5cm}
\begin{tabular}{ccccccccccccc} 

\hline\hline
\hfil& 
$c_{\sigma}^2$& $c_{\omega}^2$& $c_{\rho}^2$& $b$& $c$ &$\nu(0)$&$k_n(0)$&$x_p(0)$&$M$&$R$&$\eta_0(0)$ \\ \hline
GL& 12.684&  7.148& 4.410& 0.005610& -0.006986&-2.38799&2.71&0.24&2.37 &11.09 &0.0 \\ \hline
NL3& 15.739& 10.530& 5.324& 0.002055& -0.002650&-2.33319&2.40&0.23&2.82 &13.17 &0.0 \\ \hline
\hline

\end{tabular}
\end{center}
\end{table}

\vspace{2cm}

\newpage 

\vspace{-2cm}

{\centerline{
\epsfxsize=14cm
\epsfysize=10cm
\epsffile{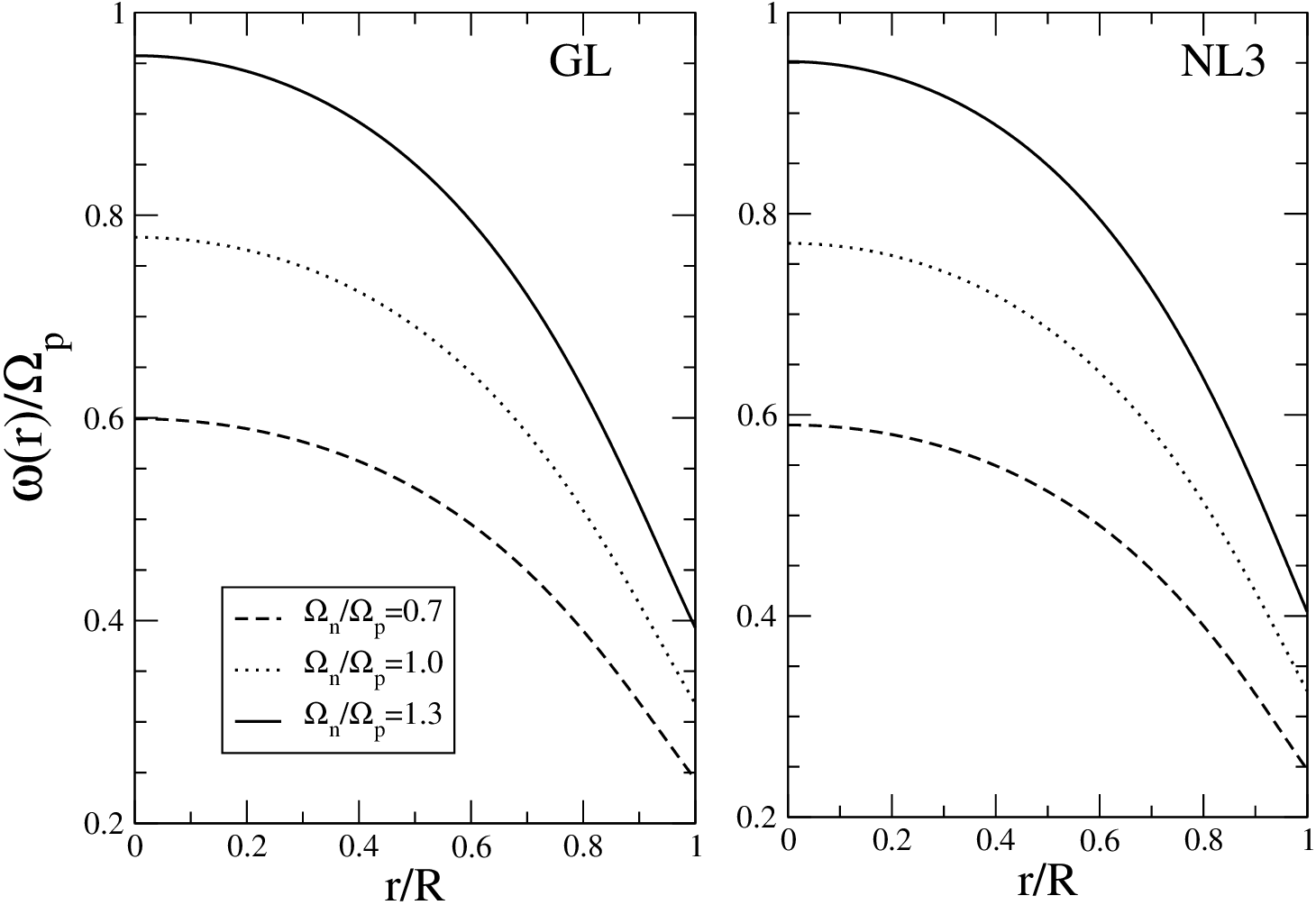}
}}
\vspace{5cm}
\noindent{\small{FIG. 1. The frame-dragging frequency $\omega(r)$ is plotted as
a function of radial distance  $(r/R)$ 
using the GL parameter set (left panel) and the NL3 parameter set (right panel)
for three different relative rotation rates $\Omega_n/\Omega_p$ .   }}
\newpage 
\vspace{5cm}

\newpage 

\vspace{-5cm}

{\centerline{
\epsfxsize=14cm
\epsfysize=10cm
\epsffile{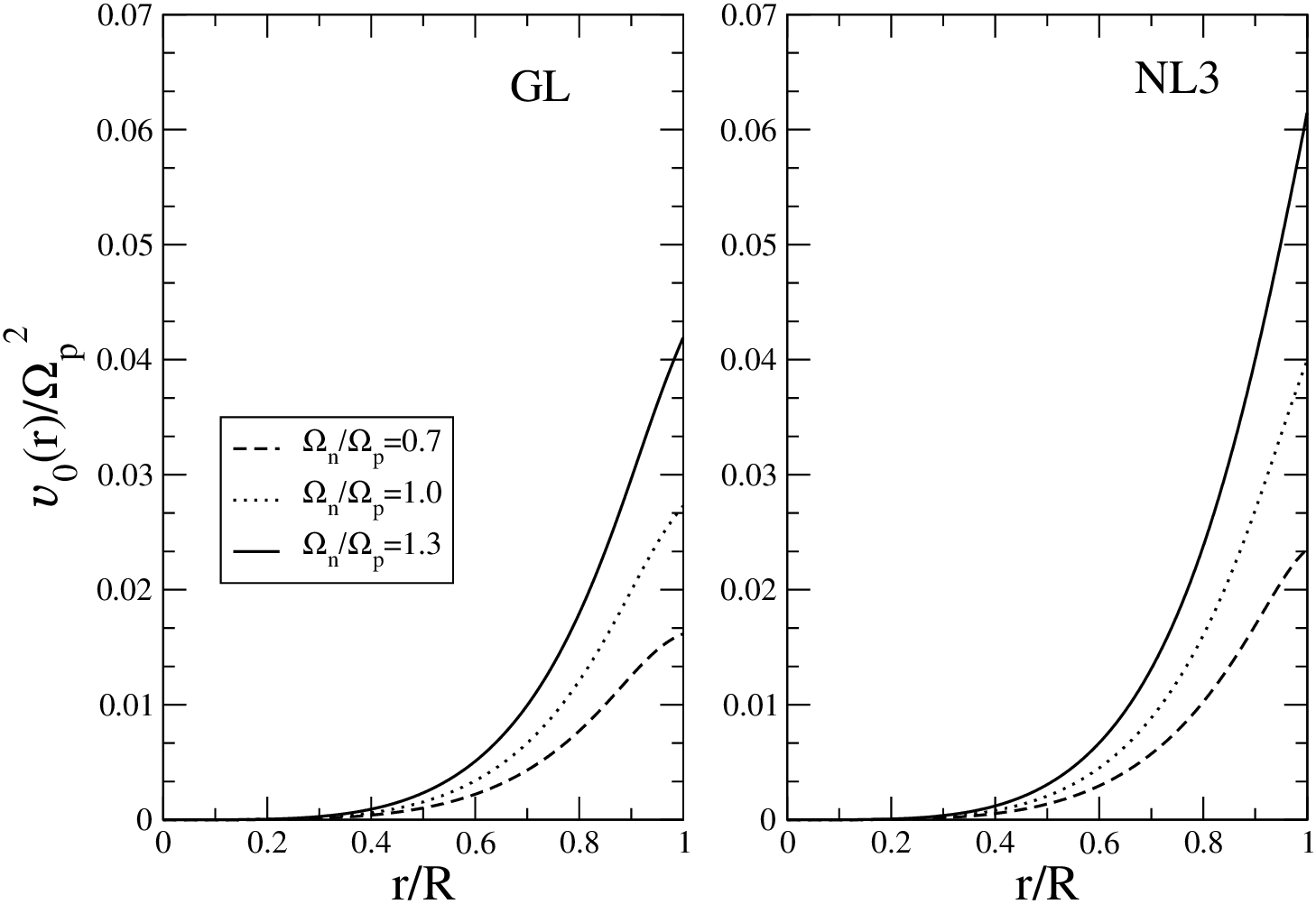}
}}
\vspace{5cm}
\noindent{\small{FIG. 2. The metric function $v_0(r)$ is  plotted as a function
of radial distance $(r/R)$ using the GL parameter set (left panel)
and the NL3 parameter set (right panel) for three different relative rotation 
rates $\Omega_n/\Omega_p$.}}

\newpage 

\vspace{-5cm}

{\centerline{
\epsfxsize=14cm
\epsfysize=10cm
\epsffile{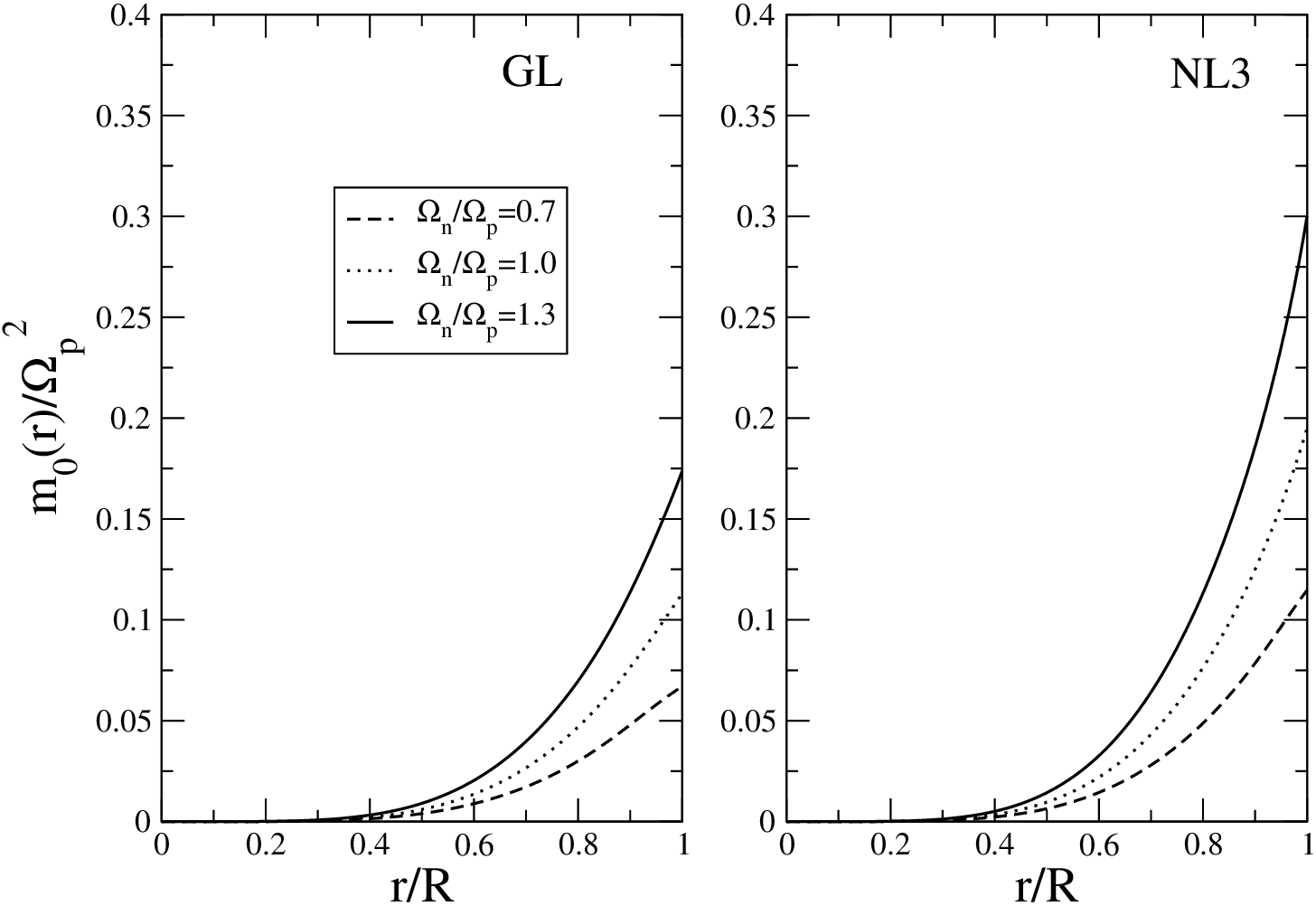}
}}
\vspace{5cm}

\noindent{\small{FIG. 3. 
The metric function $m_0(r)=r v_0(r)/\exp(\lambda(r))$ is  plotted as a function
of radial distance $(r/R)$ using the GL parameter set (left panel)
and the NL3 parameter set (right panel) for three different relative rotation 
rates $\Omega_n/\Omega_p$.}}

\newpage 

\vspace{-5cm}

{\centerline{
\epsfxsize=14cm
\epsfysize=10cm
\epsffile{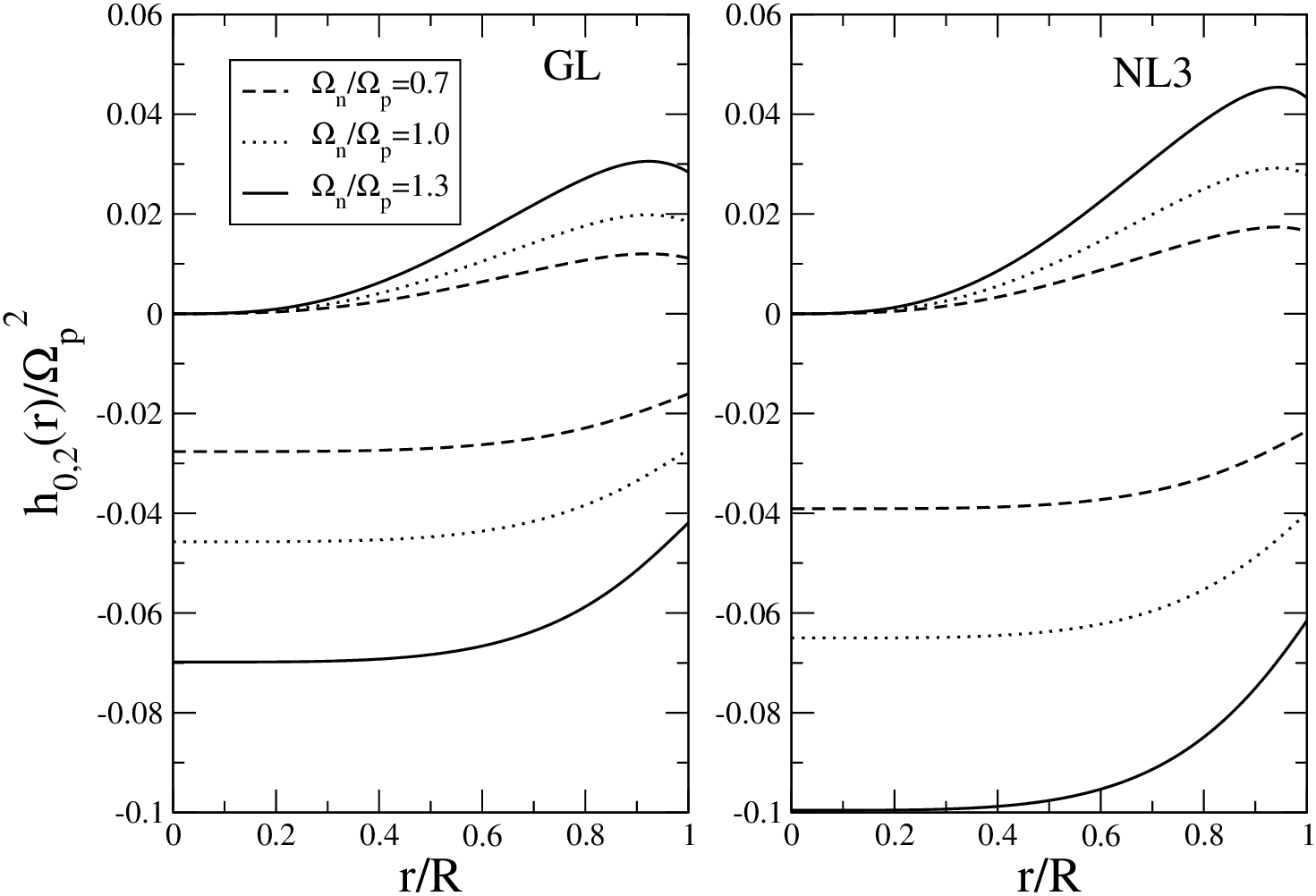}
}}

\vspace{5cm}

\noindent{\small{FIG. 4. The metric functions  $h_0(r)$ (three lower curves)  
and $h_2(r)$ (three upper curves)  are  plotted as a function 
of radial distance $(r/R)$ for three different relative rotation 
rates $\Omega_n/\Omega_p$. The left panel shows the results of the GL parameter
set and the right panel demonstrates those of the NL3 parameter set. }}

\newpage 

\vspace{-5cm}

{\centerline{
\epsfxsize=14cm
\epsfysize=10cm
\epsffile{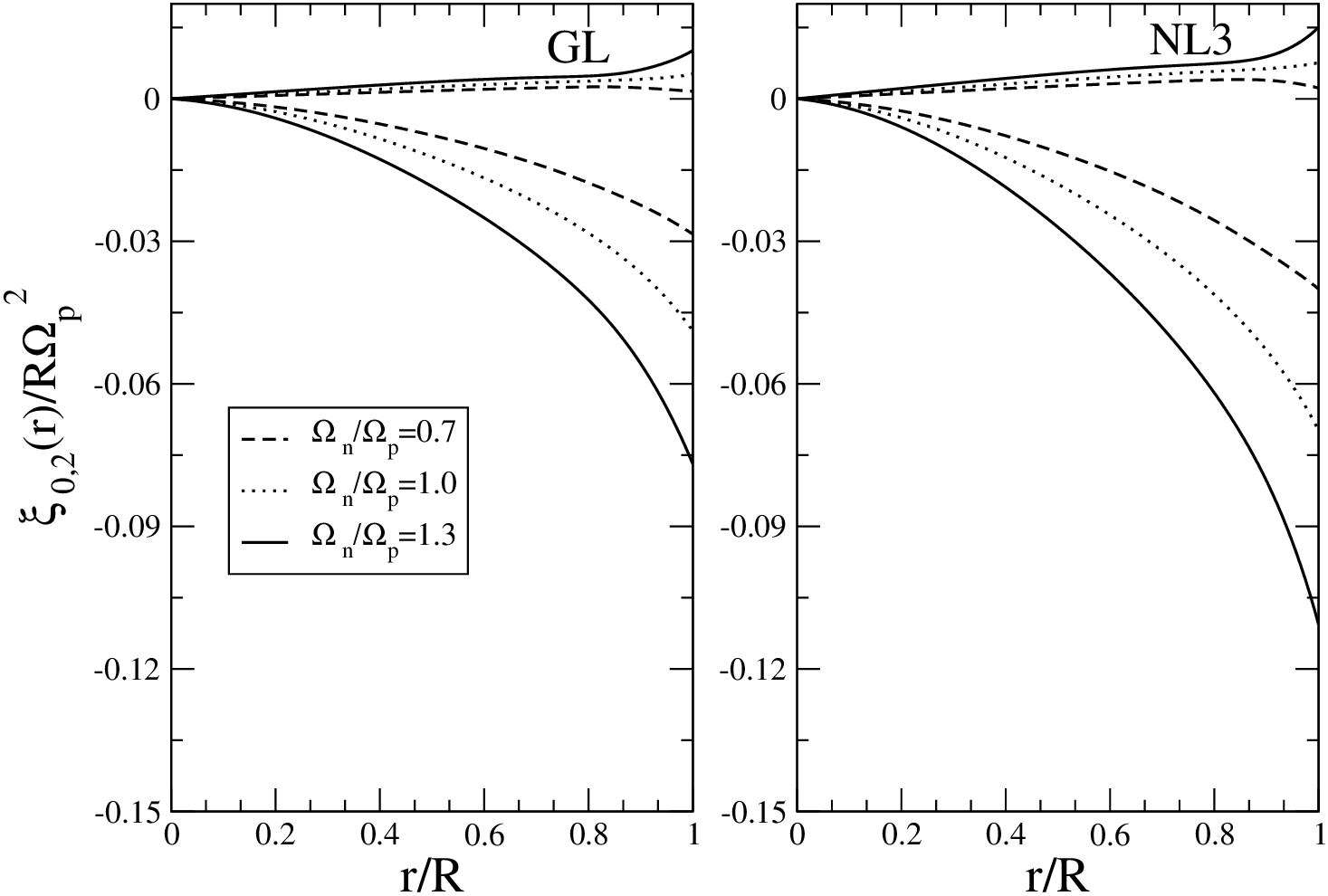}
}}
\vspace{5cm}
\noindent{\small{FIG. 5. The radial displacement $\xi_0(r)$ 
(three upper curves)  and $\xi_2(r)$ (three lower curves) are plotted as a 
function of radial distance $(r/R)$ for three different relative 
rotation rates $\Omega_n/\Omega_p$.   
The left panel shows the results of the GL parameter
set and the right panel represents those of the NL3 parameter set. }}
\newpage 

\vspace{-5cm}

{\centerline{
\epsfxsize=14cm
\epsfysize=10cm
\epsffile{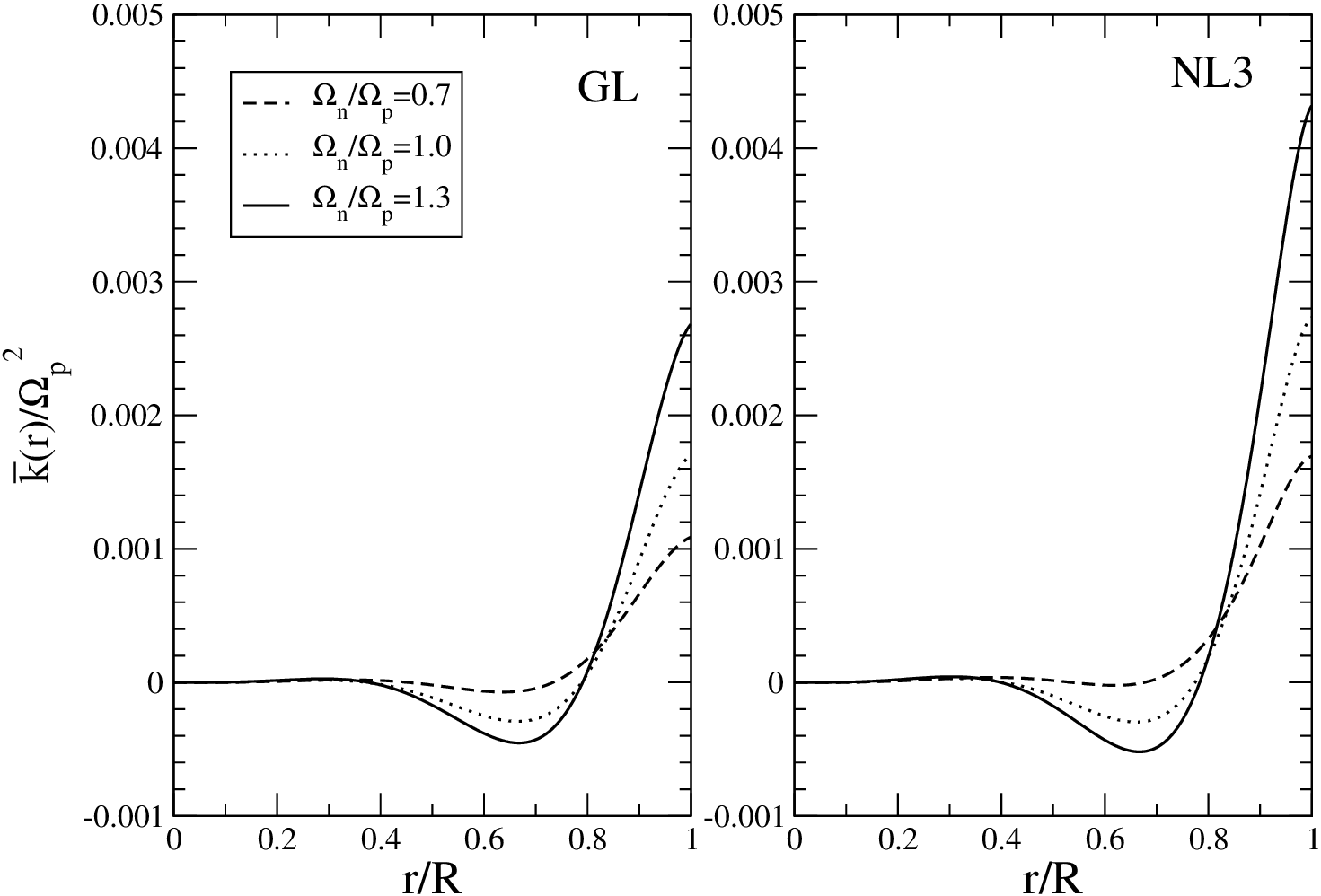}
}}

\vspace{5cm}
\noindent{\small{FIG. 6. The quantity ${\bar k}$ is plotted as a 
function of radial distance $(r/R)$ using the GL parameter set (left panel) and 
NL3 parameter set (right panel) for three different rotation rates 
$\Omega_n/\Omega_p$.}}
\newpage 

\vspace{-5cm}

{\centerline{
\epsfxsize=14cm
\epsfysize=10cm
\epsffile{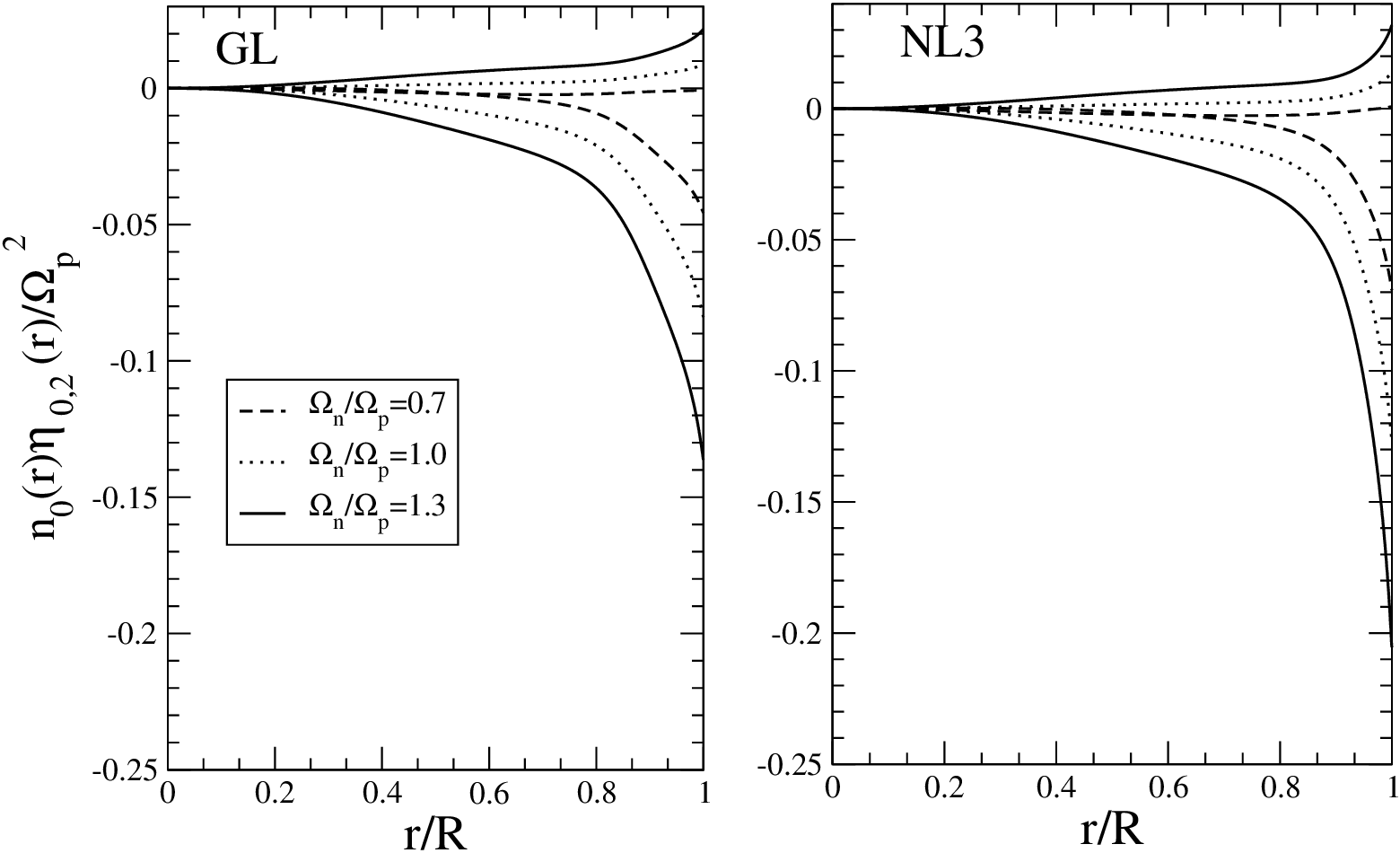}
}}
\vspace{5cm}
\noindent{\small{FIG. 7. 
The rotationally induced corrections to the 
neutron number density 
$n_0(r) \eta_0(r)$ (three upper curves) and $n_0(r) \eta_2(r)$ (three lower 
curves) are plotted as a function of radial distance $(r/R)$ for 
three different relative rotation rates $\Omega_n/\Omega_p$. Results of the GL
and NL3 parameter sets are shown in the left and right panels, respectively.}}
\newpage 

\vspace{-5cm}

{\centerline{
\epsfxsize=14cm
\epsfysize=10cm
\epsffile{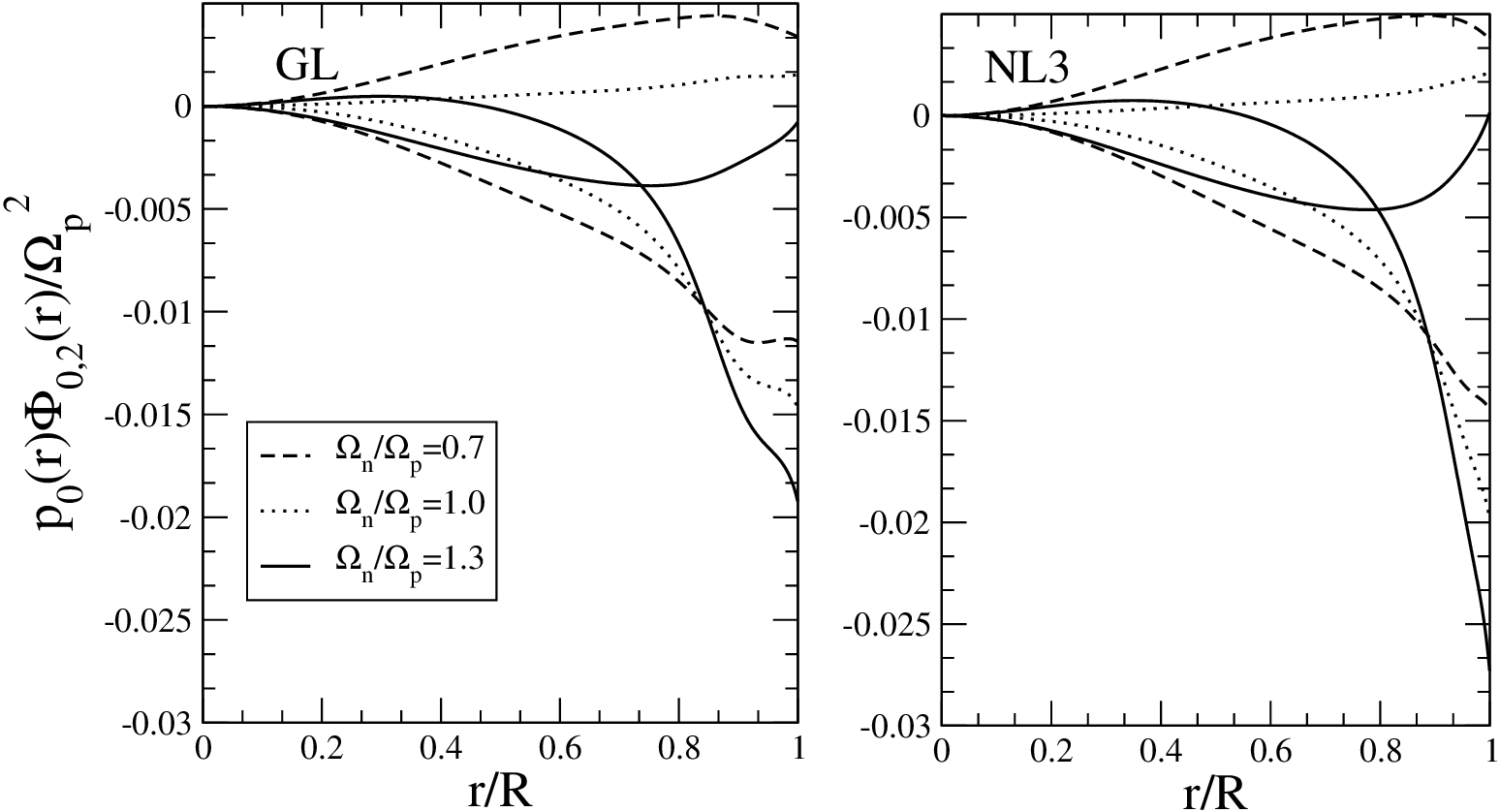}
}}
\vspace{5cm}
\noindent{\small{FIG. 8. 
The rotationally induced corrections to the 
proton number density 
$p_0(r) \phi_0(r)$ (three upper curves at $r/R=1$) and $p_0(r) \phi_2(r)$
(three lower curves at $r/R=1$) are plotted as a function of radial 
distance $(r/R)$ using the GL parameter set (left panel) and NL3 parameter set 
(right panel) for
three different relative rotation rates $\Omega_n/\Omega_p$.}}
\newpage 

\vspace{-5cm}

{\centerline{
\epsfxsize=14cm
\epsfysize=10cm
\epsffile{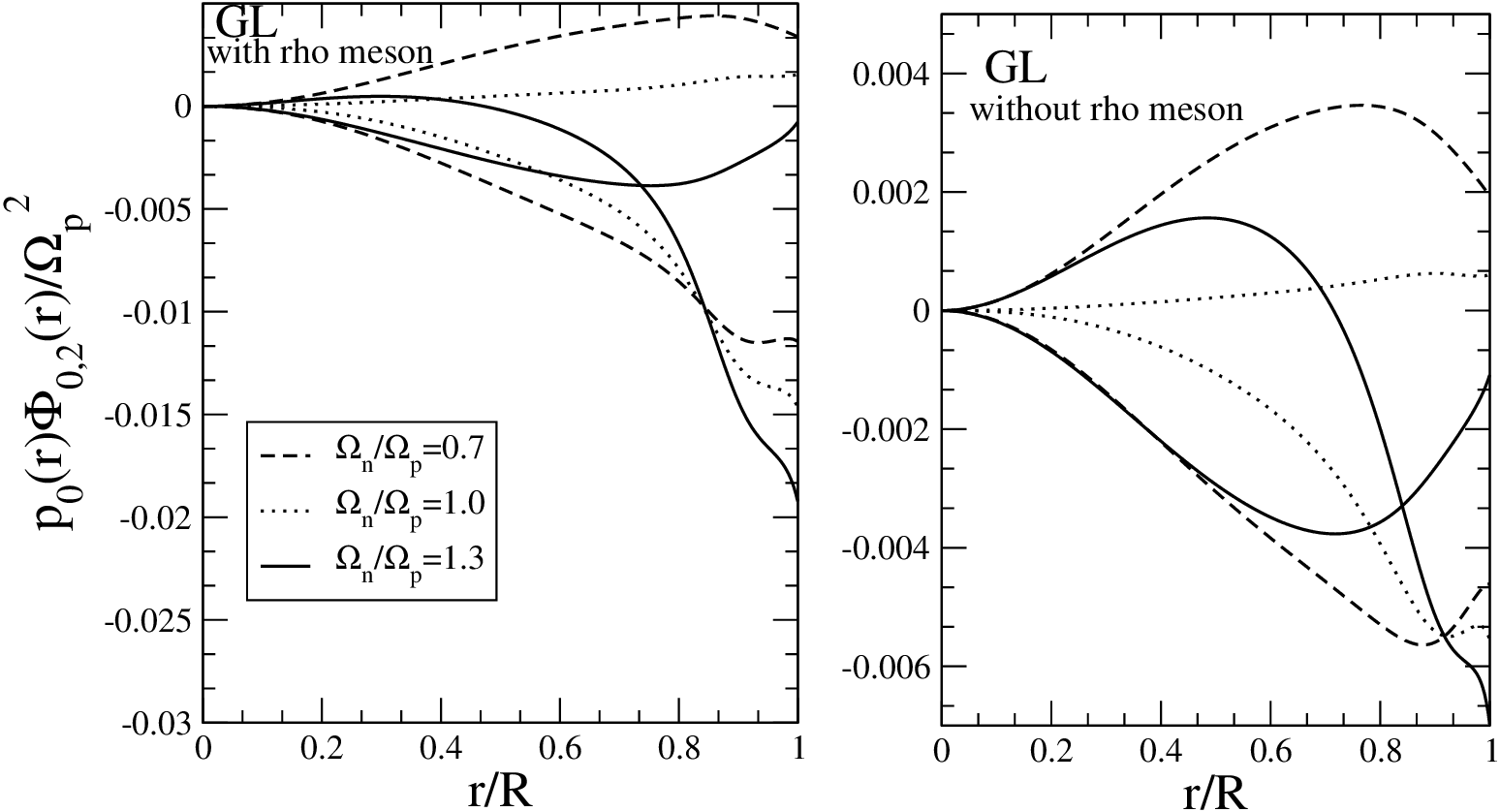}
}}
\vspace{5cm}
\noindent{\small{FIG. 9. 
The rotationally induced corrections to the 
proton number density 
$p_0(r) \phi_0(r)$ (three upper curves at $r/R=1$) and $p_0(r) \phi_2(r)$
(three lower curves at $r/R=1$) are plotted as a function of radial 
distance $(r/R)$ for the GL parameter set with (left panel) and without 
(right panel) $\rho$ mesons for three different relative rotation rates 
$\Omega_n/\Omega_p$.}}
\newpage 

\vspace{-5cm}

{\centerline{
\epsfxsize=14cm
\epsfysize=10cm
\epsffile{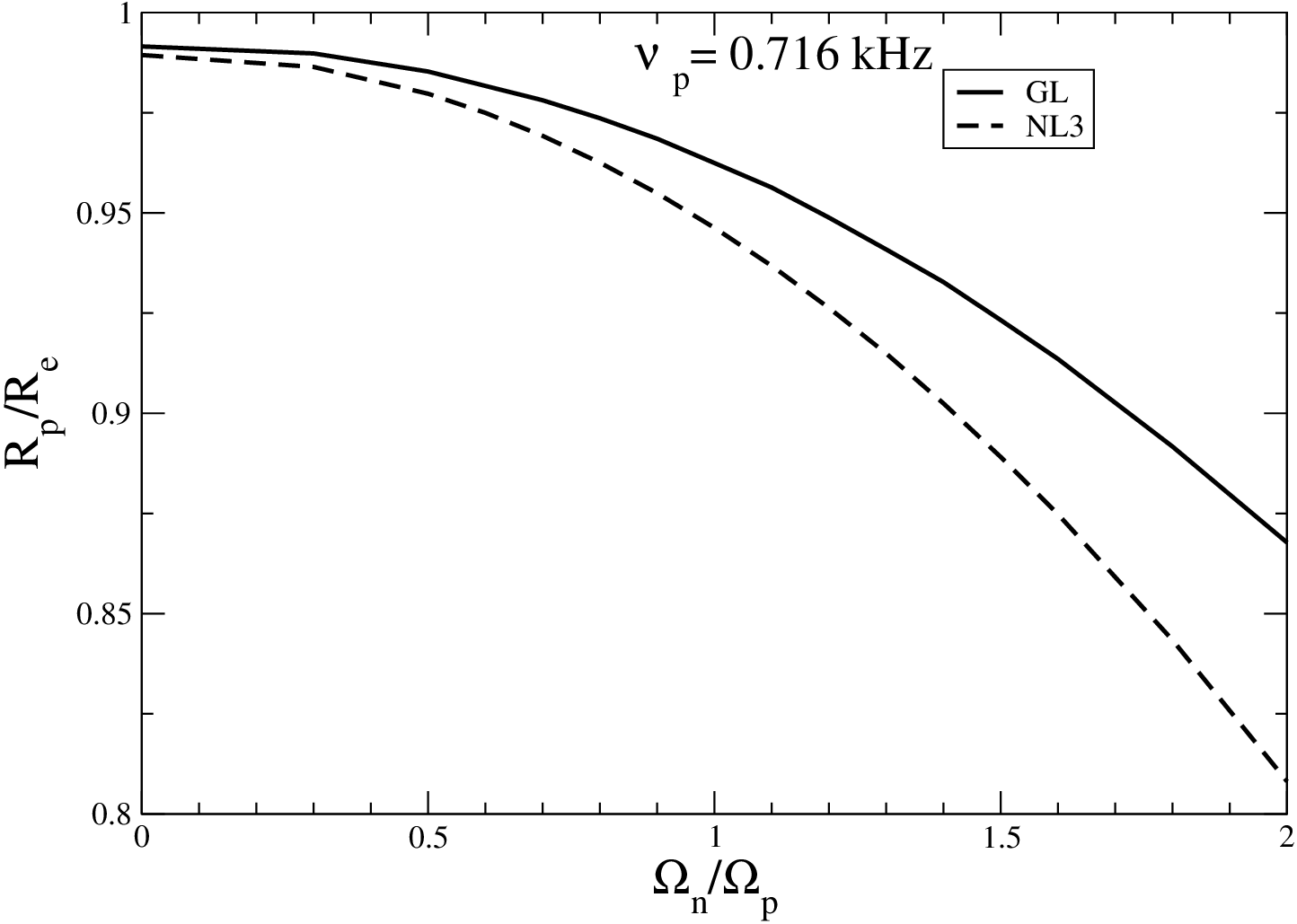}
}}
\vspace{5cm}
\noindent{\small{FIG. 10. The ratio of the polar to equatorial radii $(R_p/R_e)$
is shown as a function of relative rotation rate $\Omega_n/\Omega_p$ 
corresponding to neutron stars of masses 2.37 M$_{\odot}$ with the GL set
(solid line) and 2.82 M$_{\odot}$ with the NL3 set (dashed line), respectively,
considering that $\nu_p = \Omega_p/2\pi$ is equal to that of the fastest 
rotating pulsar having spin frequency 716 Hz \cite{hess}.}}
\newpage 

\vspace{-2cm}

{\centerline{
\epsfxsize=14cm
\epsfysize=10cm
\epsffile{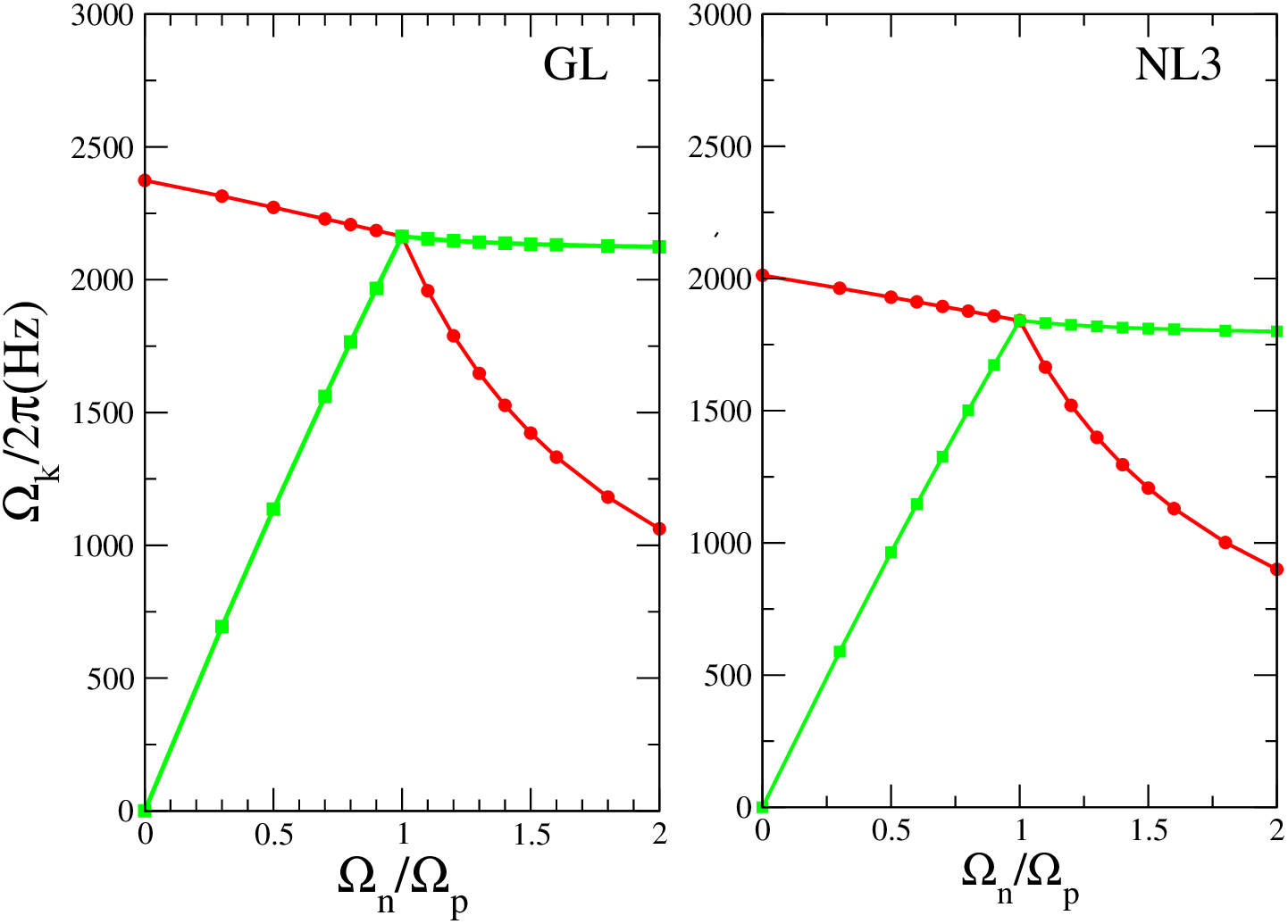}
}}
\vspace{3.0cm}

\noindent{\small{FIG. 11. The mass-shedding (Kepler) limit is 
shown as a function of relative rotation rate $\Omega_n/\Omega_p$ for the GL 
set (left panel) and the NL3 set (right panel). 
The solid squares (green) show the allowed 
rotation rate of the neutron$ fluid (\Omega_n)$ and the solid circles 
(red) show the
allowed rotation rate of the proton$ fluid (\Omega_p)$. The Kepler frequency is
the largest of the two.}}
\newpage 

\vspace{-5cm}

{\centerline{
\epsfxsize=14cm
\epsfysize=10cm
\epsffile{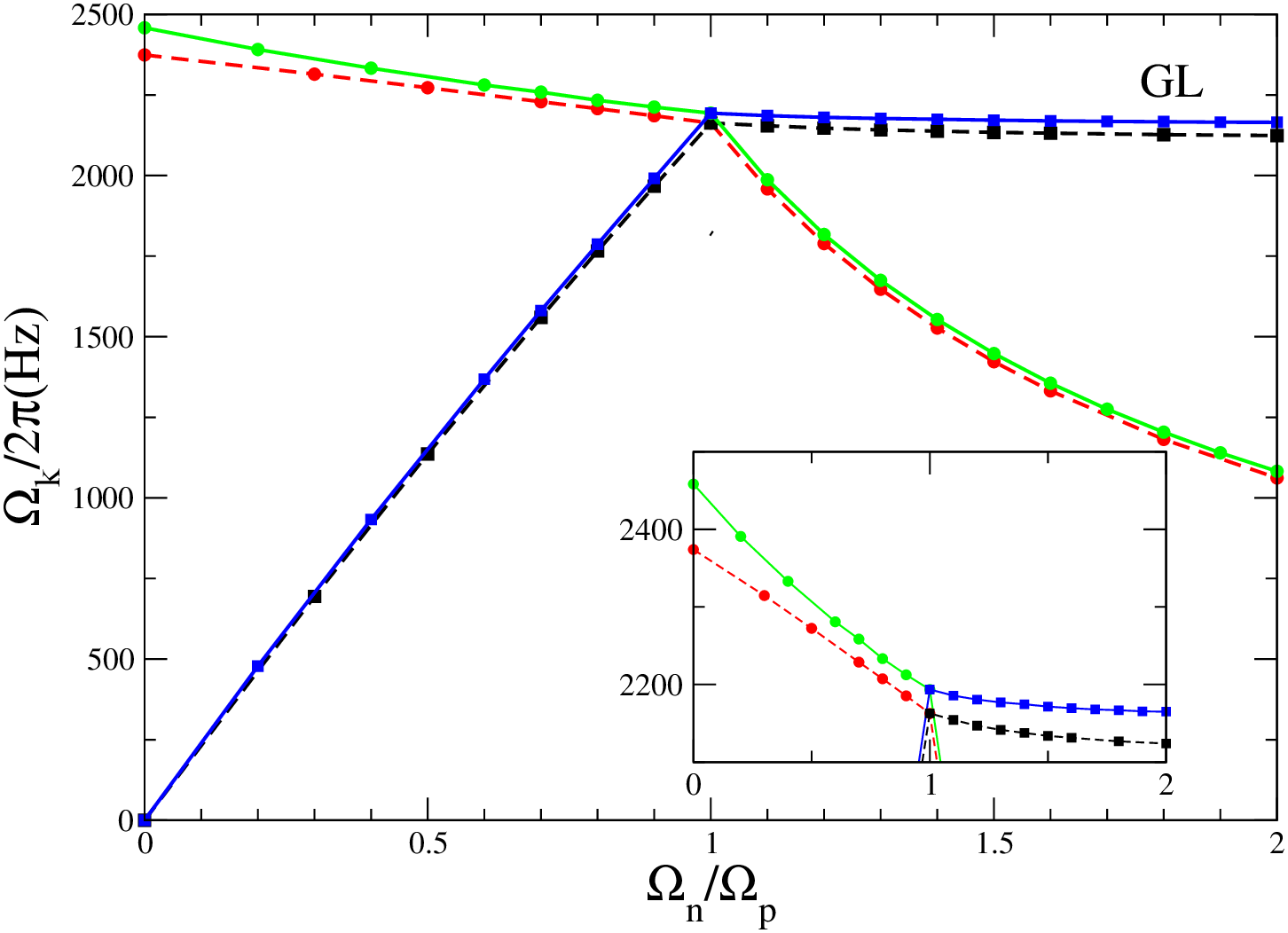}
}}
\vspace{5cm}

\noindent{\small{FIG. 12. (Color online) The mass-shedding (Kepler) limit is 
shown as a 
function of relative rotation rate $\Omega_n/\Omega_p$ with (dashed line) and 
without $\rho$ mesons (solid line) for the GL set. The solid squares show the 
allowed rotation rate of the neutron fluid $(\Omega_n)$ and the solid circles 
show the allowed rotation rate of the proton fluid $(\Omega_p)$.}}
\end{document}